\documentstyle[floats,psfig,prc,aps,manuscript]{revtex}

\begin{document}

\title{ Quenching of resonance production in heavy-ion collisions\\
        at $1\div2$ A GeV 
        \footnote{Supported by GSI Darmstadt} }

\author{ A.B. Larionov \footnote{On leave from RRC "I.V. Kurchatov Institute", 
         123182 Moscow, Russia}, W. Cassing, S. Leupold, U. Mosel }

\address{ Institut f\"ur Theoretische Physik, Universit\"at Giessen,
          D-35392 Giessen, Germany }

\maketitle

\begin{abstract}
We study the pion production in heavy-ion collisions at SIS energies within 
the coupled resonance BUU model. The standard calculation overpredicts the 
pion multiplicity in the central Au+Au collision at 1 A GeV by about a factor
of 2, which is interpreted as a signature for the quenching of the resonance 
production/absorption in nucleon-nucleon collisions at high nuclear density. 
Results of the calculations taking into account a specific form
of quenching are presented for pion spectra from light and heavy systems
as well as for the pion in-plane and out-of-plane flow.
\end{abstract}

%\vspace{0.5cm}
% 
%\hspace{-\parindent}
%PACS numbers: 25.75.-q; 25.75.Dw; 21.65.+f; 24.10.Jv; 25.75.Ld
%
%\hspace{-\parindent}
%{\it Keywords}: Au+Au at 1 A GeV, pion production, baryon resonances,
%nuclear matter, collective flow 

\section{ Introduction }

Modifications of hadrons in dense nuclear matter are currently in the 
center of theoretical studies \cite{Ko97,CB99}. In particular, the multiple 
pion production in heavy-ion collisions offers an interesting possibility
to look at the properties of the baryon resonances in nuclear matter
\cite{Eh93,Bass95,KoLi96,Teis97,Hjort97,WFN98,Eskef98,Pelte99,Lar00}.
It is well established (c.f. \cite{Eh93,Bass95}) that at beam 
energies of $1\div2$ A GeV pions are produced mostly in the two-step
process $NN \rightarrow NR$, $R \rightarrow N\pi$, where $R$ stands
for the $\Delta(1232)$ or higher baryon resonances. However, there is an 
overall tendency of transport models explicitly propagating resonances to 
overpredict the pion multiplicity in heavy-ion collisions at the beam energy 
$\sim 1$ A GeV (c.f. \cite{Pelte97}). Furthermore, taking into account 
in-medium pion dispersion relations even increases the pion yield 
\cite{Ehe93,HR97,HR98}. 

In the present work we study the effect of possible in-medium 
modifications of the reaction rates
\begin{equation}
                      N N \leftrightarrow N R              \label{NNNR}
\end{equation}
on pion production. A description of the experimental data on the pion
multiplicities requires an {\it in-medium reduction} of the rates 
(\ref{NNNR}). We propose, therefore, a density-dependent modification
of the transition rates (\ref{NNNR}), which gives a reasonable description 
of the pion multiplicities in the systems C+C, Ni+Ni and Au+Au at the beam 
energy of 1$\div$2 A GeV. We will show, moreover, that this modification
is consistent with other pion observables, i.e. transverse mass and kinetic
energy spectra, angular distributions and in-plane collective flow.

The structure of the paper is as follows: In Sect. II we discuss the effect 
of the resonance production/absorption quenching on the pion production. 
Sect. III contains the results on the collective pion flows, while in Sect. 
IV the summary and conclusions are given.

\section{ Density dependent transition rates $N N \leftrightarrow N R$ }

In our numerical calculations we have used the BUU model in the version
described in Ref. \cite{EBM99}. The nucleons, $\Delta(1232)$ and higher
(up to $D_{35}(2350)$) baryon resonances, $\Lambda$ and $\Sigma$ 
hyperons, mesons $\pi,~\eta,~\omega,~K,~\rho,~\sigma$ are explicitly
propagated in this BUU implementation. The model has been successfully 
applied to the calculation of nucleon collective flow observables in 
heavy-ion collisions at energies from $\sim100$ AMeV up to $\sim2$ A GeV
\cite{LCGM00}. A good description of "more elementary" $\gamma$ and
$e^-$ induced reactions has been also reached within the model
(see Refs. \cite{EBM99,Lehr00}).

In this work we concentrate on pion production in nucleus-nucleus 
collisions continuing the studies initiated in Refs. \cite{Teis97,Lar00}.
All calculations in the present paper are performed using the soft 
momentum-dependent mean field (SM) with incompressibility $K=220$ MeV, which 
gives the best overall description of the nucleon collective flow variables
\cite{LCGM00}. The BUU set of parameters, which includes the SM mean field 
as well as vacuum resonance parameters and vacuum elastic and inelastic cross 
sections, will be denoted as "standard" below. The modifications of the 
standard set will be discussed in the text explicitly. 

First, we compare the calculated pion multiplicities in central Au+Au 
collisions at 1 A GeV with the experimental data. In Fig. 1 the dashed line
shows the pion number vs time in the case of the standard calculation. The 
pion multiplicity asymptotically reaches a value of $\sim60$ which is almost
a factor of 2 larger than the experimental value of $\sim34$ (c.f. Ref. 
\cite{Pelte97}). Pions are produced dominantly by the baryon resonance 
decays. Therefore, the pion multiplicity is very sensitive to the number 
of resonances present in the system. The differential production cross section
of a resonance $R$ with mass $M$ in a nucleon-nucleon collision, averaged over 
the spins of initial particles and summed over the spins of final particles,
is given as
\begin{equation}
{d \sigma_{N_1 N_2 \rightarrow N_3 R_4} \over d M^2 d \Omega } =
 {1 \over 64 \pi^2} \overline{|{\cal M}|^2} 
{p_{34} \over p_{12} \varepsilon^2} {\cal A}(M^2) 
\times 2(2J_R+1)~,                             \label{signnnr}
\end{equation}
where $\overline{|{\cal M}|^2}$ is a spin-averaged matrix element squared,
$p_{12}$ and $p_{34}$ are c.m. momenta of incoming and outgoing particles,
$\varepsilon$ is the total c.m. energy, ${\cal A}(M^2)$ is the spectral
function of the resonance, and $J_R$ is the spin of the resonance. 
The inverse reaction cross section is
\begin{equation}
{d \sigma_{N_3 R_4 \rightarrow N_1 N_2} \over d \Omega} =
 {1 \over 64 \pi^2} \overline{|{\cal M}|^2} 
{p_{12} \over p_{34} \varepsilon^2} \times {4 \over C_{12}}~,
                                               \label{signrnn}
\end{equation}
where, due to detailed balance, the same matrix element appears as in
Eq. (\ref{signnnr}), $C_{12}=2$ if the nucleons $N_1$ and $N_2$ are 
identical and $C_{12}=1$ otherwise.  

The matrix element which appears in Eqs. (\ref{signnnr}),(\ref{signrnn})
should be evaluated in nuclear matter.
Instead of using any model dependent parametrizations for the in-medium
matrix element, we adopt a simple relation
\begin{equation}
  \overline{|{\cal M}|^2} 
  = \kappa(\rho) \overline{|{\cal M}_{vac}|^2}~,~~~~~~
    \kappa(0)=1~,                                        \label{muinmed}
\end{equation}
where $\overline{|{\cal M}_{vac}|^2}$ is the vacuum matrix element squared
and $\kappa(\rho)$ is some density-dependent function to be determined
by comparison with the experimental data. The in-medium matrix elements
for the non-resonant (background) one-pion production/absorption
$N_1 N_2 \leftrightarrow N_3 N_4 \pi$, double $\Delta$(1232) 
production/absorption $N_1 N_2 \leftrightarrow \Delta_3 \Delta_4$
and resonance scattering on nucleons $N_1 R_2 \rightarrow N_3 R_4$ are
also related to their vacuum values by Eq. (\ref{muinmed}) with the same 
universal function $\kappa(\rho)$.

We address now the pion multiplicity in central Au+Au collisions
at 1.06 A GeV using various choices of the function $\kappa(\rho)$.
As a first trial, we simulate an enhancement of resonance production
in NN collisions \cite{BBKL88} assuming a linear dependence on density as 
in Refs. \cite{Lar00,Engel94}:
\begin{equation}
          \kappa(\rho) = 1 + \beta {\rho \over \rho_0}~,~~~~
                         \beta > 0~,
                                                    \label{kappa}
\end{equation}
where $\rho_0=0.16$ fm$^{-3}$ is the equilibrium nuclear matter density. 
We see from Fig. 1 (dotted line), that the amplified matrix element with 
$\kappa(\rho) = 1 + 3\rho/\rho_0$ leads to an about 20\% reduction in
the final pion multiplicity which, however, is still above the experimental 
abundance. This decrease originates from the amplified -- by detailed balance 
-- absorption of resonances in the expansion stage of the system. A further 
increase of the parameter $\beta$ practically does not reduce the pion 
number any further (see Fig. 3 for $\beta > 0$). 

An essential reduction of the pion multiplicity can only be reached by
a {\it quenching} of the resonance production/absorption processes
in nuclear matter at high density. To this aim we apply a piece-wise
linear function of the form
\begin{equation}
   \kappa(\rho) 
 = \min(1,\max(0,1 + \beta (\rho/\rho_0-1)))~,~~~~
                         \beta < 0~,
                                                    \label{kappafit}
\end{equation}
which is shown by the solid line in Fig. 2 for the particular case 
$\beta=-2$. Within the parametrization (\ref{kappafit}), the in-medium 
effects vanish at subnuclear densities ($\kappa=1$), but above $\rho_0$ 
the nuclear medium creates the shadowing of binary collisions involving 
resonances leading, effectively, to a quenching of these collisions for
$\rho \geq (1-1/\beta)\rho_0$.
In Fig. 3 (see the part of the BUU curve for $\beta < 0$) we report 
the number of produced negative pions vs $\beta$ in the case of the 
parametrization (\ref{kappafit}). The proper pion multiplicity is 
reproduced for $\beta=-2$; we will denote this calculation as ``quenched''
in the following. For this calculation, resonances do not experience any 
elastic or inelastic scatterings in high density nuclear matter. They can, 
however, decay or be produced in the processes $R \leftrightarrow N \pi$. 
In the quenching scenario, the resonance production in NN collisions happens 
only at relatively low density $\rho \leq 1.5\rho_0$, i.e. dominantly in 
the overlapping tails of the density profiles of colliding nuclei. 
This reduces the maximum number of $\Delta$-resonances to about $35\%$
as compared to a standard calculation (c.f. solid and dashed lines in 
the lower panel of Fig. 1).

Before discussing other pion-related observables we check, how the resonance
quenching influences the underlying nucleon dynamics. Fig. 4 shows the proton 
rapidity distribution and the transverse momentum of protons in the reaction 
plane as a function of rapidity. Due to a lower nucleon inelasticity within 
the quenching scenario, the nuclei show less stopping, i.e. the rapidity 
distribution becomes somewhat wider and, since also the pressure is decreased, 
a slightly smaller transverse in-plane flow is generated making the agreement 
of the standard SM calculation \cite{LCGM00} with the proton flow data worse 
(see also Fig. 11). This problem, however, can be resolved by either 
increasing the incompressibility or the stiffness of the momentum-dependent 
interaction. The latter modifications influence only weakly the pion 
multiplicity and, therefore, are not studied in this work.

In Fig. 5 we present the pion multiplicity as a function of the participant
number $A_{part}$ for Au+Au collisions at 1.06 A GeV in comparison with the
FOPI and TAPS data collected in Ref. \cite{Pelte97}. We have determined 
the participant number for a given impact parameter using the 
participant-spectator model with sharp nuclear density profiles. 
The quenched calculation is in a good agreement with the 
data for central collisions; in peripheral collisions the medium effects are 
smaller, the difference between the standard and the quenched calculation 
vanishes gradually and, thus, the multiplicity is still somewhat 
overpredicted. This discrepancy in peripheral collisions could arise from 
different definitions of the participant number $A_{part}$ in our 
calculations and in Ref. \cite{Pelte97}, since the impact parameter can
not be measured directly in the experiment. 

Fig. 6 shows the beam energy dependence of the charged pion multiplicity 
for central Ni+Ni collisions compared with the FOPI data from Ref.
\cite{Pelt97}. The quenched calculation describes the data rather well. 
With increasing beam energy the difference between standard and quenched 
calculations reduces, since the nonresonant string mechanism for pion 
production becomes dominant, which is the same in both calculations.
We recall that in our transport approach strings are excited in NN
collisions for invariant collision energies $\sqrt{s} \geq 2.6$ GeV
corresponding to the beam energy of $1.73$ A GeV if the Fermi
motion is neglected. 

Fig. 7 shows the c.m. kinetic energy spectra of charged pions at 
$\Theta_{c.m.}=130^o$ for Au+Au at 1.06 A GeV. The quenching improves 
the agreement with the data somewhat producing a more concave shape 
of the spectrum for $E_{kin}^{c.m.} > 0.1$ GeV. However, the calculated 
cross sections in this region are still too large, while at low kinetic
energy the experimental data are underpredicted by the quenched 
calculation.

A similar tendency is visible in the case of $p_t$-spectra at midrapidity 
shown in Fig. 8: the improvement by the quenching mechanism at high $p_t$ 
is evident, but at low $p_t$ too few pions are produced. This leads also
to the underprediction of the midrapidity pion yield by the quenched 
calculation. Therefore, since the total pion multiplicity agrees with
the data (or slightly overpredicts for peripheral collisions, see Fig. 5),
the quenched calculation will produce some excess of pions at projectile/%
target rapidities. We will, however, discard here the comparison of our 
results with the FOPI data on the pion rapidity distribution from Ref. 
\cite{Pelte97}. This comparison would require the treatment of our  
BUU-events in exactly the same way as the data, which is out of scope of 
this work.

The reason for the underprediction of the low-$p_t$ pion yields by the 
quenched calculation is, probably, an overdamping of the low mass 
resonance production in NN collisions at high density by the quenching 
factor (\ref{kappafit}). Indeed, a smaller medium effect is expected for 
higher kinetic energies of the final particles in analogy to Pauli blocking 
\cite{Com1}. In this exploratory work we are not aiming at a detailed 
description of the pion energy spectra, but, rather, want to demonstrate 
the gross effect of the density modified matrix element on various pion 
observables. It is straightforward, however, to introduce a $\sqrt{s}$ and 
resonance-mass-dependent modification factor $\kappa$ in Eq.(\ref{muinmed}).
Another physical effect is the off-shellness of the pion in the nuclear medium 
due to large $\pi N \rightarrow \Delta$ cross section, which leads to a broad
pion spectral function. Thus a dynamical evolution of the hadron spectral
functions in BUU will modify the low momentum spectra sensitively \cite{LCLM}.

In Fig. 9 we present the c.m. polar angle distributions of charged pions.
The shape of these spectra is similar to the one for free-space pion 
production in NN collisions. There is a good overall agreement of 
the quenched calculation with the data.

For the collisions C+C at 0.8$\div$2 A GeV (Fig. 10) the calculated
$m_t$-spectra of $\pi^0$'s are practically identical in the standard and
quenched calculations. This is expected, since the in-medium modifications
are smaller in the lighter system.

\section{ Collective flow }

In earlier studies (c.f. \cite{Bass95,BALi94}), the pion antiflow and
squeeze-out have been explained by the rescattering of pions
on the spectator nucleons, which leads to the shadowing phenomenon.

Fig. 11 shows the proton, $\pi^+$ and $\pi^-$ in-plane flow
($:= d<p_x>/dY^{(0)}$ at $Y^{(0)}=0$) for the Au+Au system at 1.15 A GeV 
as a function of impact parameter in comparison to the data \cite{Kint97}. 
The best overall agreement with the data is reached within the standard 
SM calculation. For $\pi^\pm$, the quenched and standard results are almost 
indistinguishable within statistics. However, the proton flow is 
underpredicted by the quenched calculation (see discussion in the previous 
section). The difference in flows for $\pi^+$ and $\pi^-$ is due to the 
Coulomb interaction with protons which increases (decreases) the antiflow 
for $\pi^+$ ($\pi^-$) for large impact parameter. 

In Fig. 12 we show the squeeze-out ratio 
$R_N := (N_{\pi^+}(90^o)+N_{\pi^+}(-90^o))/(N_{\pi^+}(0^o)+N_{\pi^+}(180^o))$ 
as a function of the transverse momentum for positive pions at midrapidity
in comparison to the data \cite{Shin98}.
The squeeze-out ratio grows with $p_t$ since faster moving pions in 
transverse direction reach the spectators earlier and rescatter in 
analogy to the case of the nucleon squeeze-out \cite{LCGM00}.
A peculiar feature of the $R_N(p_t)$-dependence for positive pions is its 
saturation at relatively low $p_t \sim 0.3$ GeV/c, while the nucleon
squeeze-out ratios increase monotonically with $p_t$ 
(c.f. \cite{Lambr94,Brill96}). This behaviour of the pion squeeze-out
ratio is related to the resonance nature of $\pi N$ scattering
\cite{BALi94}. Our calculations, both standard and quenched, underpredict 
the absolute value of $R_N$, but the shape of the function $R_N(p_t)$ is in 
a qualitative agreement with the data. The reason for this discrepancy lies, 
most probably, in the absence of any momentum dependence in the pion potential.

\section{ Summary and conclusions }

Within the transport BUU model \cite{EBM99} we have studied the influence 
of the in-medium modifications of the reaction rates (\ref{NNNR})
on pion production in heavy-ion collisions around 1 A GeV. 
The pion multiplicity in the system Au+Au at 1 A GeV can be
consistently described assuming a quenching of the resonance
production/absorption in NN collisions at high nuclear densities. The 
quenching generally improves the description of standard BUU calculations 
with experiment on other pion observables: $p_t$-, $m_t$- and kinetic 
energy spectra, in-plane and out-of-plane collective flows.

The physical reason for the in-medium reduction of the resonance production
rates, that seems to be needed to describe the pion data, remains open.
This is in particular so because there are theoretical predictions that the
$\Delta$ excitation cross section should increase with density \cite{BBKL88}.
A collisional broadening of the $\Delta$-resonance, that was not taken into
account in Ref. \cite{BBKL88}, could, in principle, provide the obviously 
needed reduction, but the calculations of Ref. \cite{Effe97}, that were based
on reasonable collision rates, have shown that this broadening is too small.
All these models have in common that they work with free coupling constants 
and thus neglect any in-medium vertex corrections. Good calculations of 
inelastic nucleon-nucleon scattering in nuclear matter would be needed to 
clarify this problem.

\newpage
 
\section*{ Figure captions }
 
\begin{description}  

\item[Fig. 1] Pion number (upper panel) and $\Delta$-resonance number
(lower panel) vs time for a central collision Au+Au at 1.06 A GeV.
Standard calculations are represented by dashed lines. The dash-dotted 
and dotted lines show the cases of in-medium amplified reaction rates 
(\ref{NNNR}) for $\beta=1$ and $\beta=3$, respectively, while the solid
lines correspond to the quenched calculation ($\beta=-2$). The experimental 
data from Ref. \cite{Pelte97} are within the dashed band in the upper panel.

\item[Fig. 2] Density dependence of the modification factor $\kappa(\rho)$ 
of the matrix element for resonance production/absorption 
(see Eq.(\ref{muinmed})). The notation of lines is the same as in Fig. 1.

\item[Fig. 3] Number of negative pions for the central collision Au+Au
at 1.06 A GeV as a function of the coefficient $\beta$ in the density 
dependence of the factor $\kappa$ (see Eqs.(\ref{kappa}),(\ref{kappafit})).
The part at $\beta < 0$ corresponds to the parametrization (\ref{kappafit}), 
while $\kappa(\rho) = 1 + \beta\rho/\rho_0$ for $\beta > 0$. The dashed band 
represents the experimental data from Ref. \cite{Pelte97}. 

\item[Fig. 4] Proton rapidity distribution (upper panel) and rapidity
dependence of the proton transverse momentum in the reaction plane
(lower panel) for Au+Au collision at 1.06 A GeV and impact parameter b=6 fm. 
The rapidity is normalized to the projectile rapidity in the 
center-of-mass system: $Y^{(0)} \equiv (y/y_{proj})_{c.m.}$.

\item[Fig. 5] $\pi^-$, $\pi^0$ and $\pi^+$ multiplicities (left, central and 
right panels, respectively) vs participant number $A_{part}$ for the system 
Au+Au at 1.06 A GeV. The experimental data are from Ref. \cite{Pelte97}.  

\item[Fig. 6] $\pi^-$ and $\pi^+$ multiplicities (left and right panels, 
respectively) in central collisions of Ni+Ni as a function of the beam energy. 
The BUU calculations are performed for b=1 fm. 
The experimental data are from Ref. \cite{Pelt97}.

\item[Fig. 7] C.m. kinetic energy inclusive spectra of negative and positive
pions (upper and lower panels, respectively) at $\Theta_{c.m.}=130^o$
for Au+Au at 1.06 A GeV. The experimental data are from Ref. \cite{Pelte97}.

\item[Fig. 8]  $\pi^-$, $\pi^0$ and $\pi^+$ transverse momentum inclusive 
spectra at midrapidity for Au+Au at 1.06 A GeV. The spectra are calculated 
in the rapidity intervals $Y^{(0)}=-0.2\div0.2$ for $\pi^\pm$
and $Y^{(0)}=-0.25\div0.21$ for $\pi^0$, but normalized to a rapidity 
interval $dy=1$ following Refs. \cite{Pelte97,Schwalb94}. 
The experimental data are from Refs. \cite{Pelte97,Schwalb94}.

\item[Fig. 9] C.m. polar angle distributions of charged pions in 
Au+Au collisions at 1.06 A GeV. Left and right panels show,
respectively, the $\pi^-$ and $\pi^+$ distributions for the c.m.
kinetic energy $E_{kin}^{c.m.} > 40$ MeV, while the central panel
presents the $\pi^-$ distribution for $E_{kin}^{c.m.} > 365$ MeV.
The experimental data are from Ref. \cite{Pelte97}.

\item[Fig. 10] Transverse mass spectra of $\pi^0$'s in selected
rapidity intervals near midrapidity (following Ref. \cite{Aver97}) for 
C+C collisions at 0.8, 1 and 2 A GeV: $y_{lab}=0.42\div0.74,~0.42\div0.74$ 
and $0.8\div1.08$, respectively. The thick solid lines show exponential 
fits to the experimental data from Ref. \cite{Aver97}.

\item[Fig. 11] In-plane flow as a function of impact parameter for protons
(upper panel), $\pi^+$'s (middle panel) and $\pi^-$'s (lower panel) in
comparison with the experimental data from Ref. \cite{Kint97} for Au+Au at 
1.15 A GeV. 

\item[Fig. 12] Transverse momentum dependence of the squeeze-out
ratio $R_N$ for positive pions for semicentral Au+Au collisions at
1 A GeV. The calculated results are impact-parameter-weighted in the range
$b=5\div7$ fm. The experimental data are from Ref. \cite{Shin98}.

\end{description}

\clearpage
\thispagestyle{empty}
 
% Fig. 1
\begin{figure}[btp]
\psfig{figure=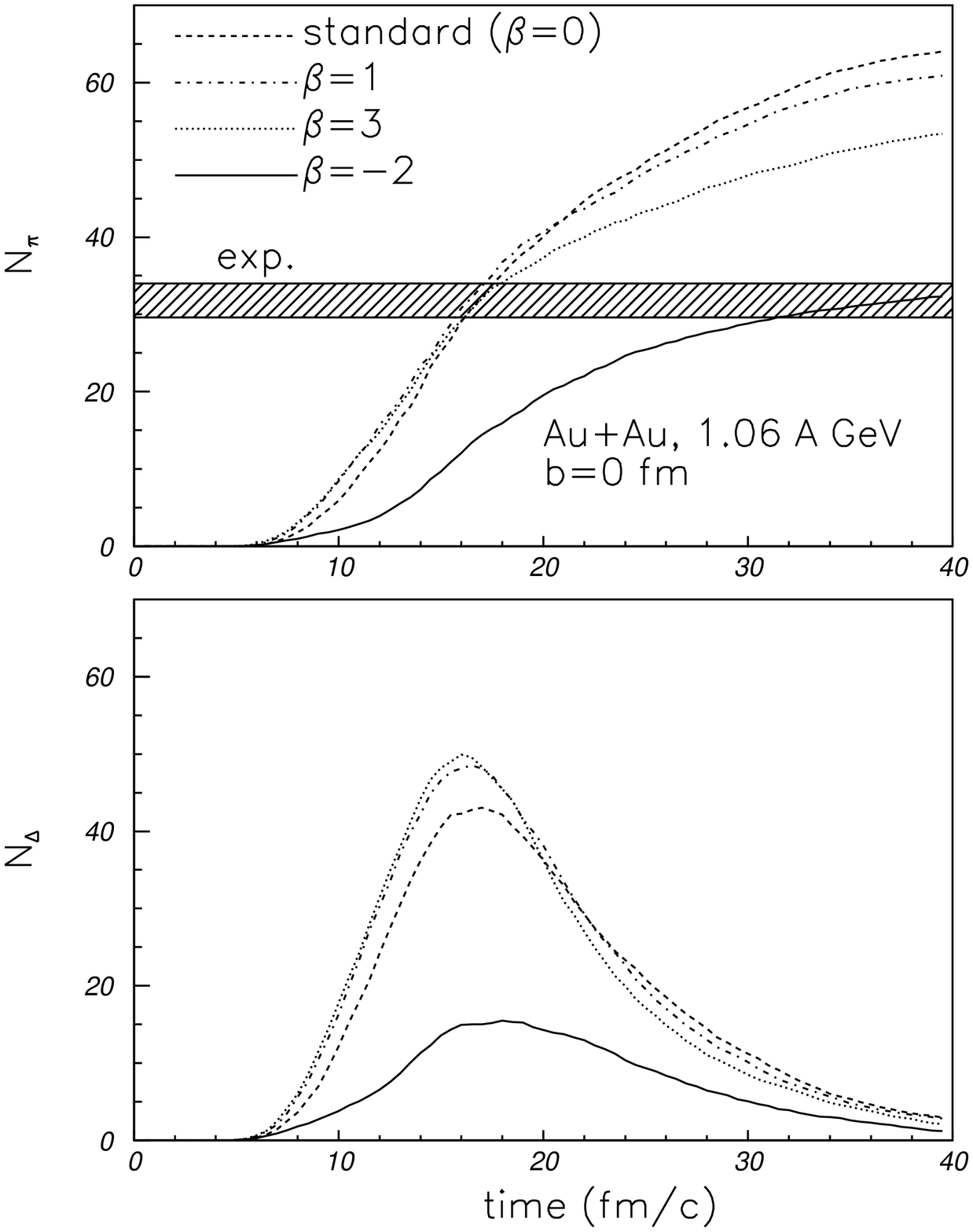,width=0.8\textwidth}
\caption{ }
\end{figure}

\clearpage
\thispagestyle{empty}

% Fig. 2
\begin{figure}[btp]
\psfig{figure=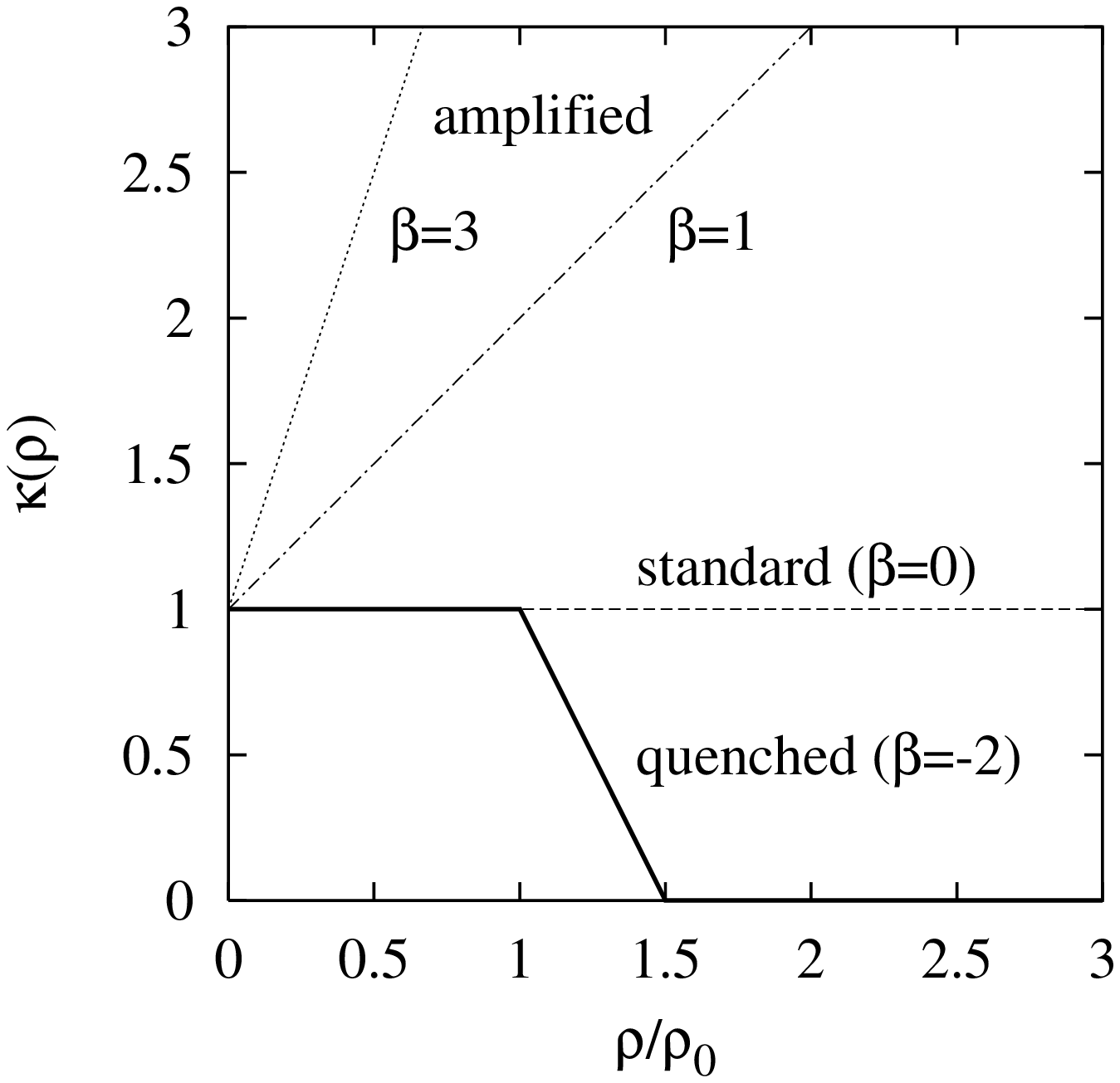,width=0.8\textwidth}

\vspace{2cm}

\caption{ }
\end{figure}

\clearpage
\thispagestyle{empty}

% Fig. 3
\begin{figure}[btp]
\psfig{figure=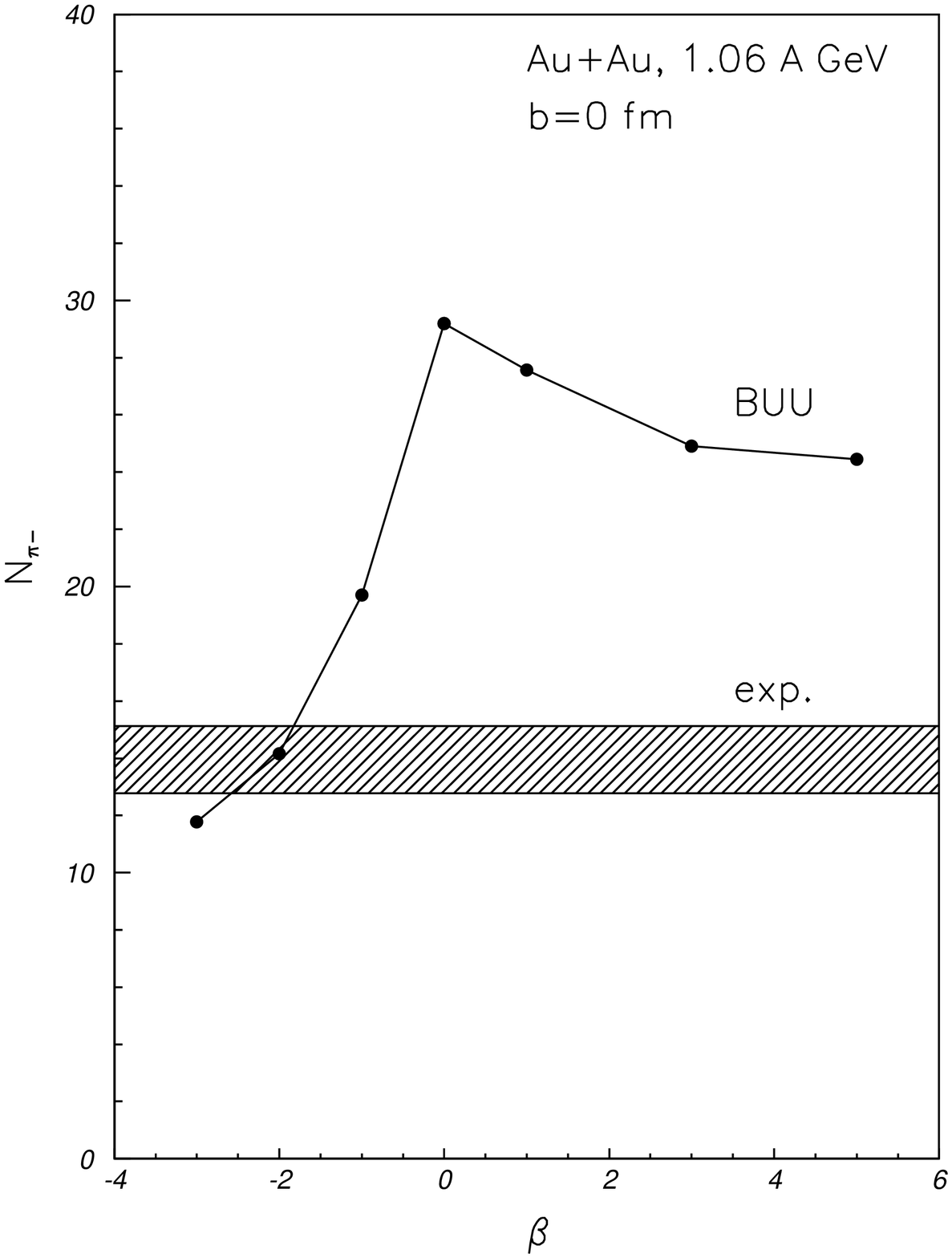,width=0.8\textwidth}
\caption{ }
\end{figure}

\clearpage
\thispagestyle{empty}

% Fig. 4
\begin{figure}[btp]
\psfig{figure=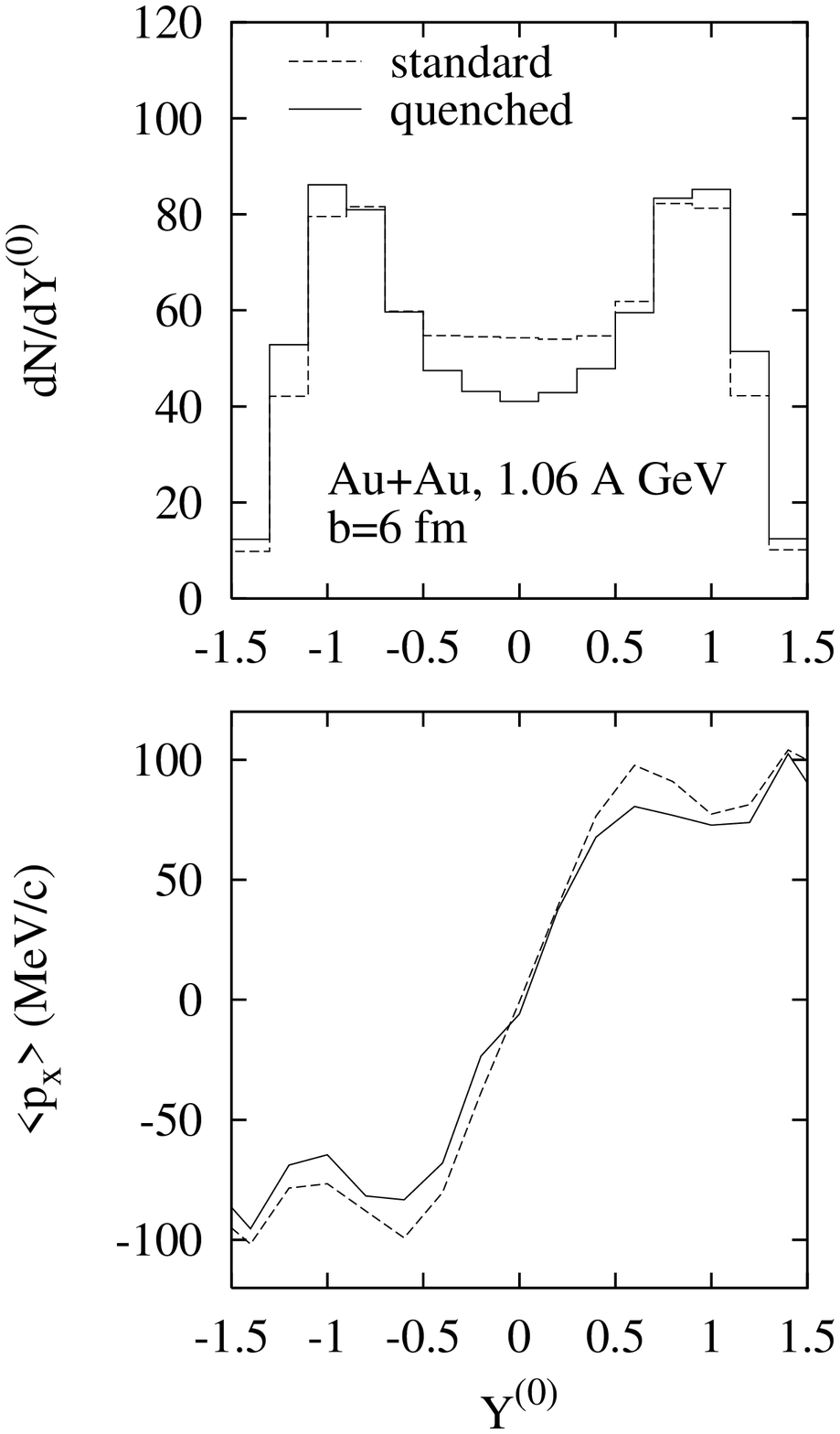,width=0.8\textwidth}

\vspace{2cm}

\caption{ }
\end{figure}

\clearpage
\thispagestyle{empty}

% Fig. 5
\begin{figure}[btp]
\psfig{figure=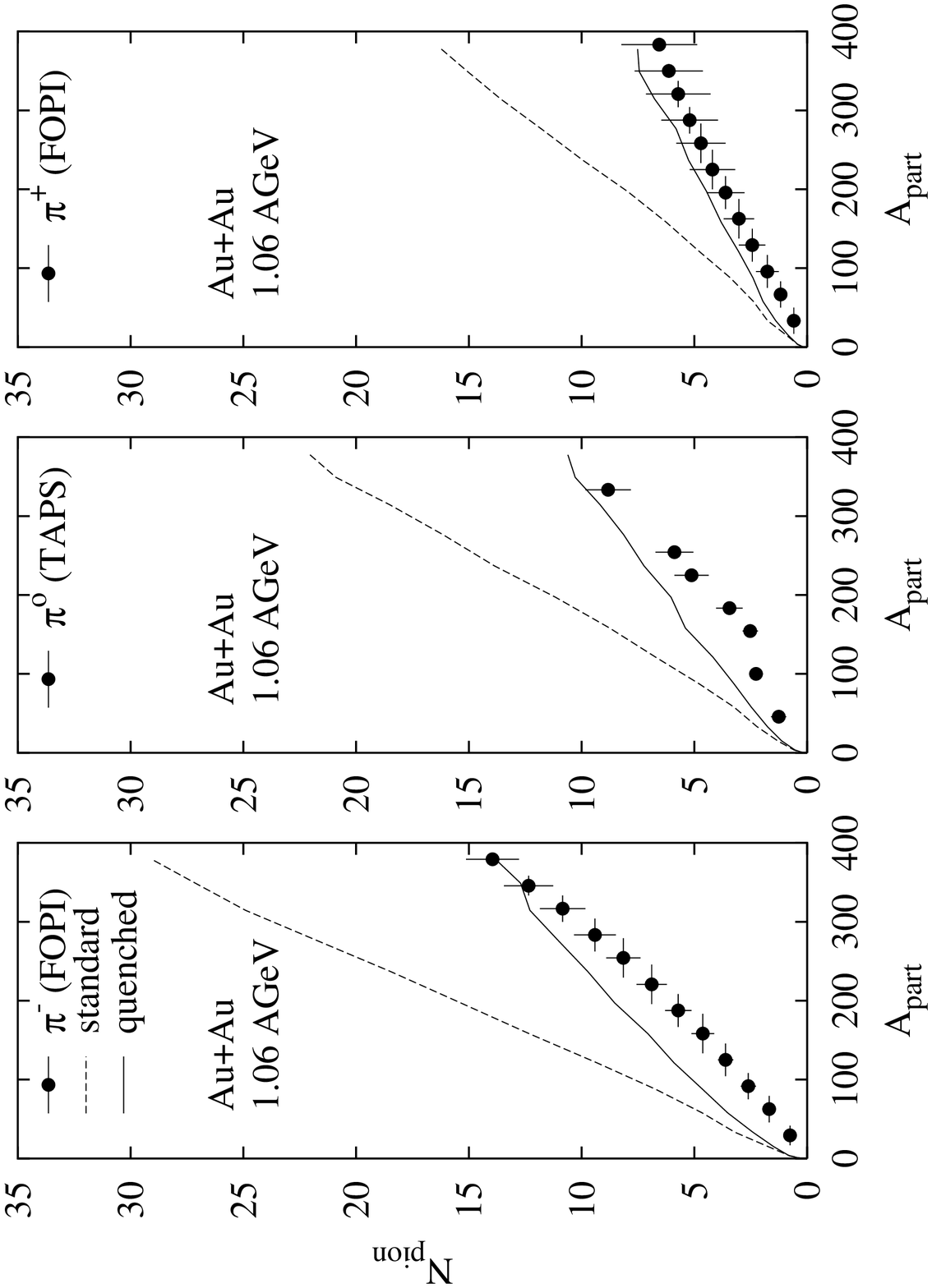,width=0.8\textwidth}

\vspace{3cm}

\caption{ }
\end{figure}

\clearpage
\thispagestyle{empty}

% Fig. 6
\begin{figure}[btp]
\psfig{figure=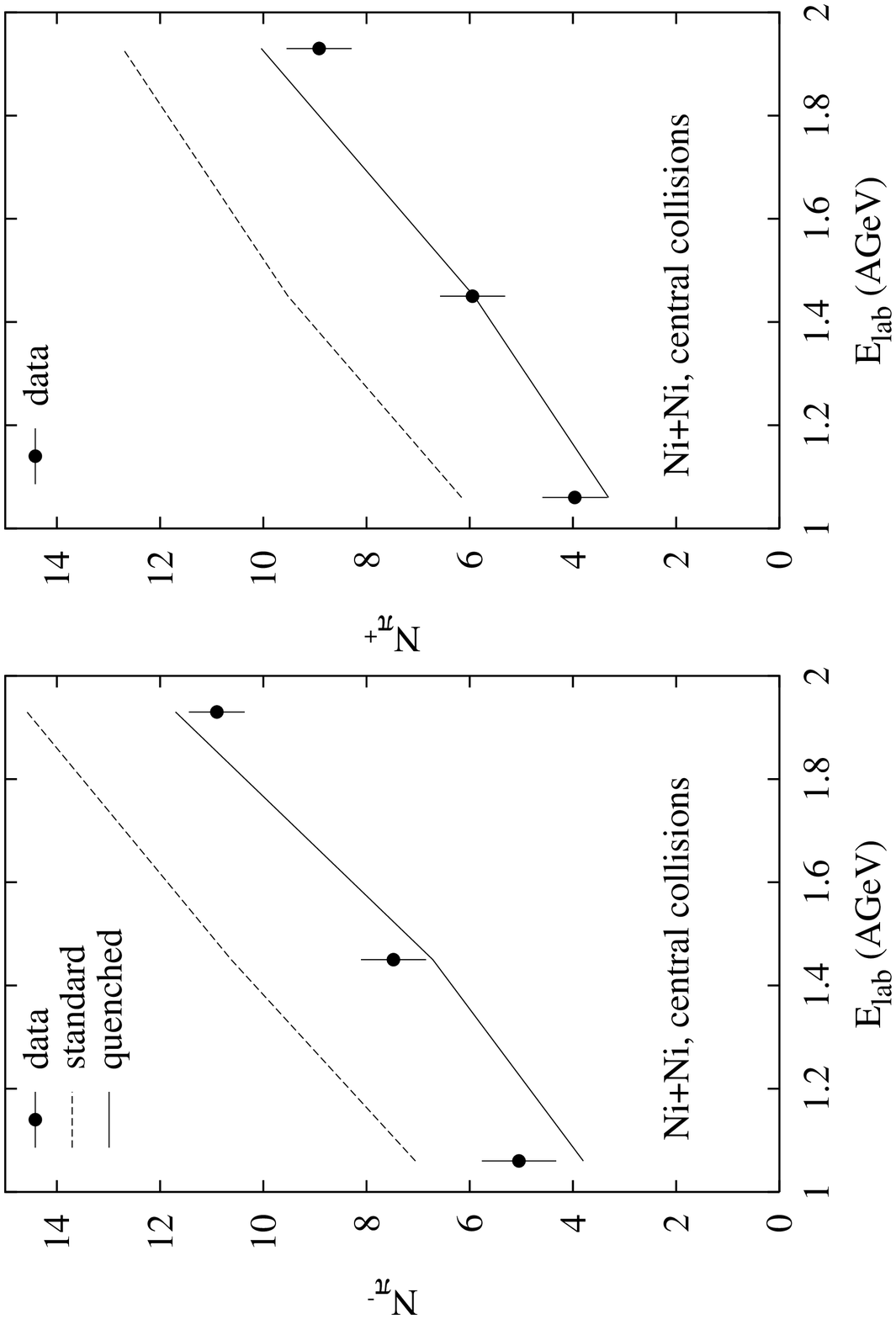,width=0.8\textwidth}

\vspace{2cm}

\caption{ }
\end{figure}

\clearpage
\thispagestyle{empty}

% Fig. 7
\begin{figure}[btp]
\psfig{figure=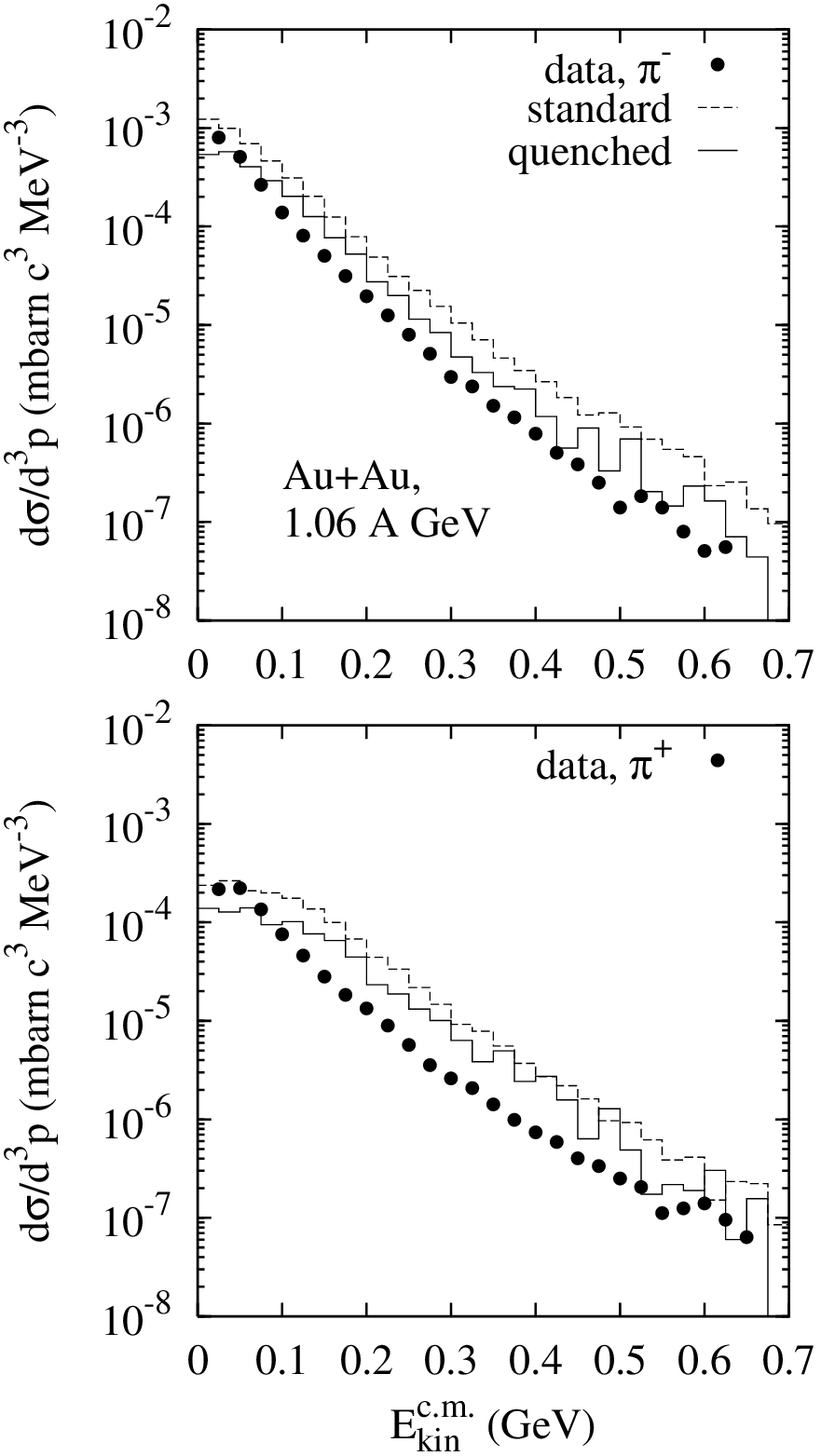,width=0.8\textwidth}

\vspace{2cm}

\caption{ }
\end{figure}

\clearpage
\thispagestyle{empty}

% Fig. 8
\begin{figure}[btp]
\psfig{figure=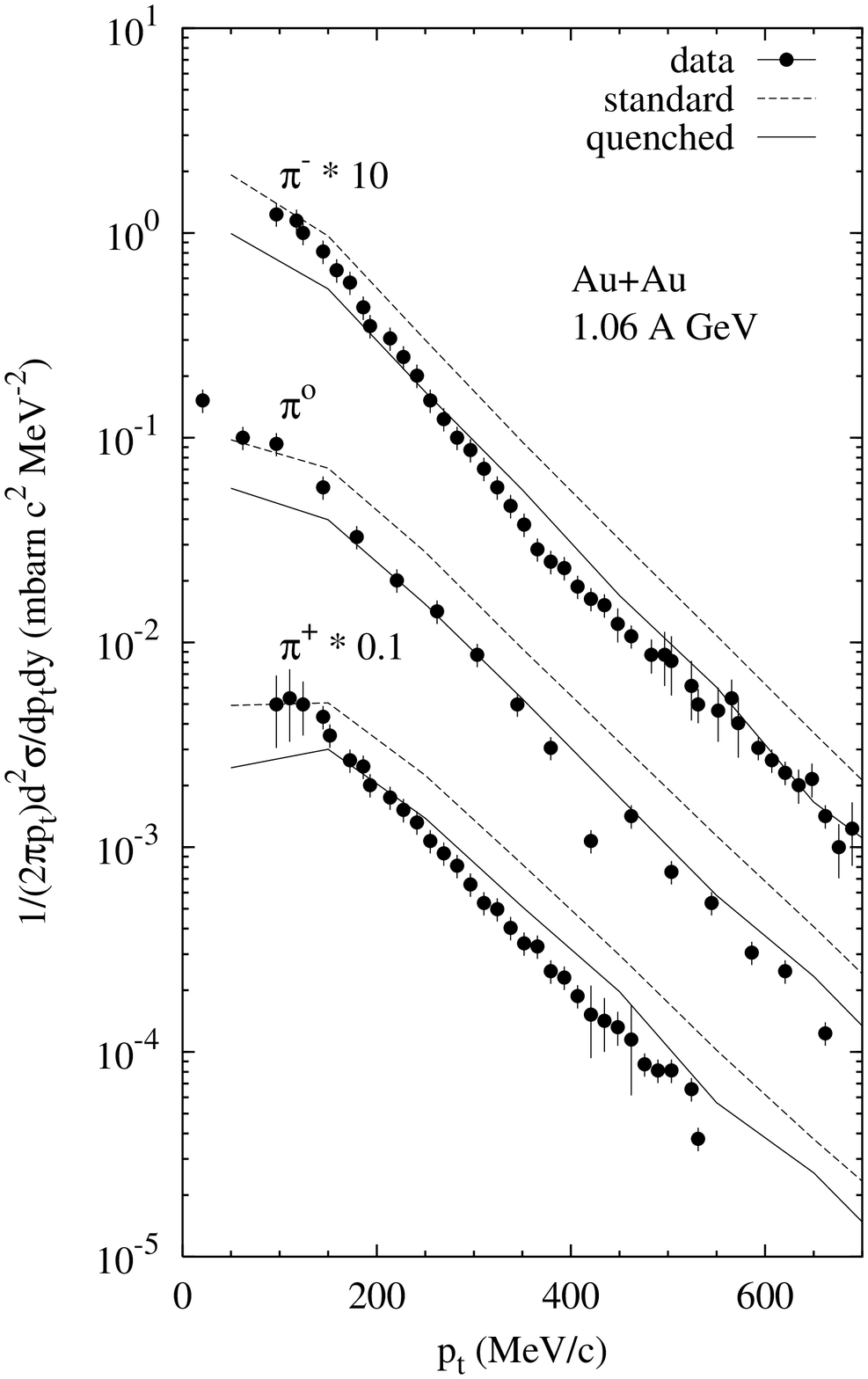,width=0.8\textwidth}

\vspace{2cm}

\caption{ }
\end{figure}

\clearpage
\thispagestyle{empty}

% Fig. 9
\begin{figure}[btp]
\psfig{figure=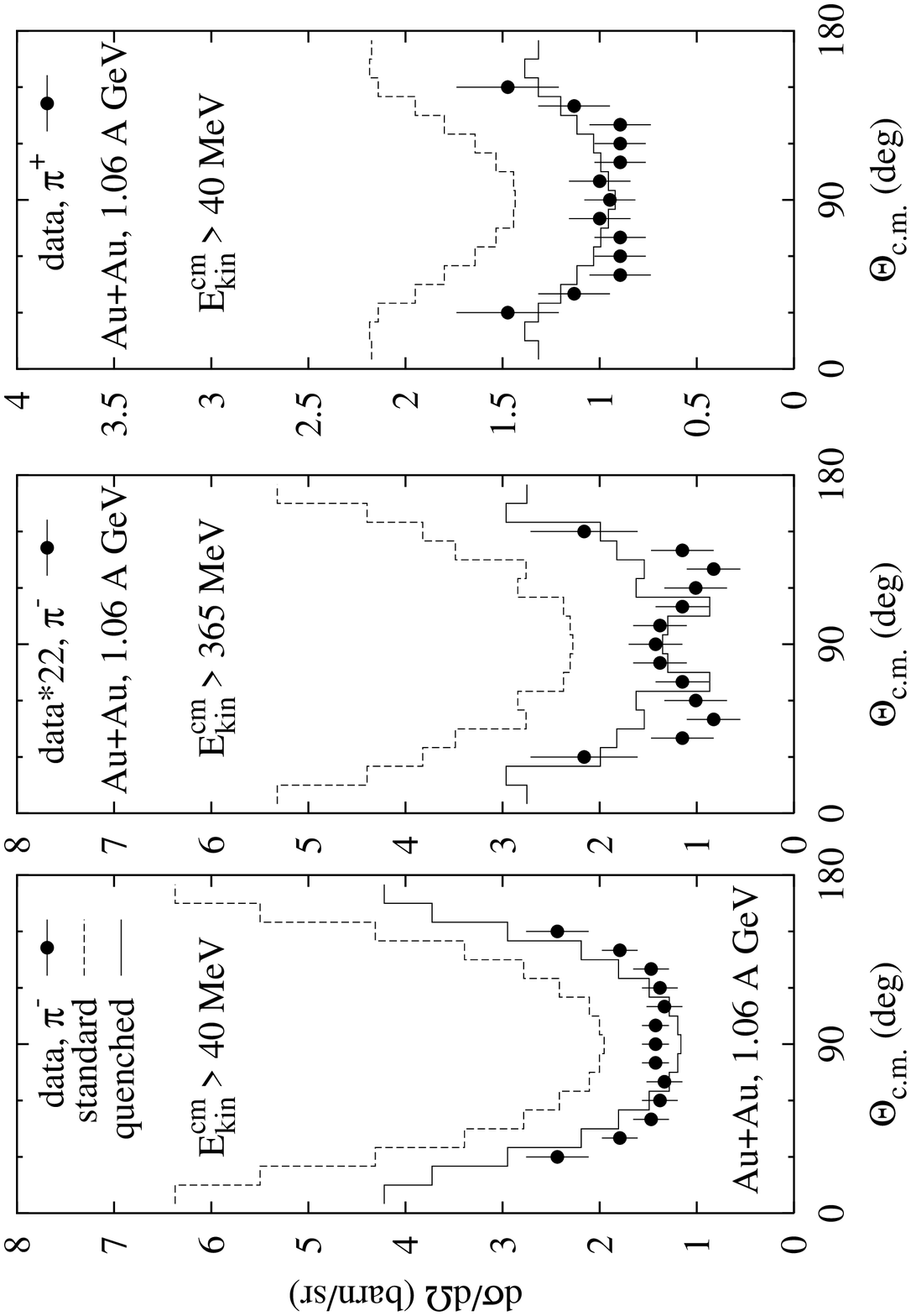,width=0.8\textwidth}

\vspace{3cm}

\caption{ }
\end{figure}

\clearpage
\thispagestyle{empty}

% Fig. 10
\begin{figure}[btp]
\psfig{figure=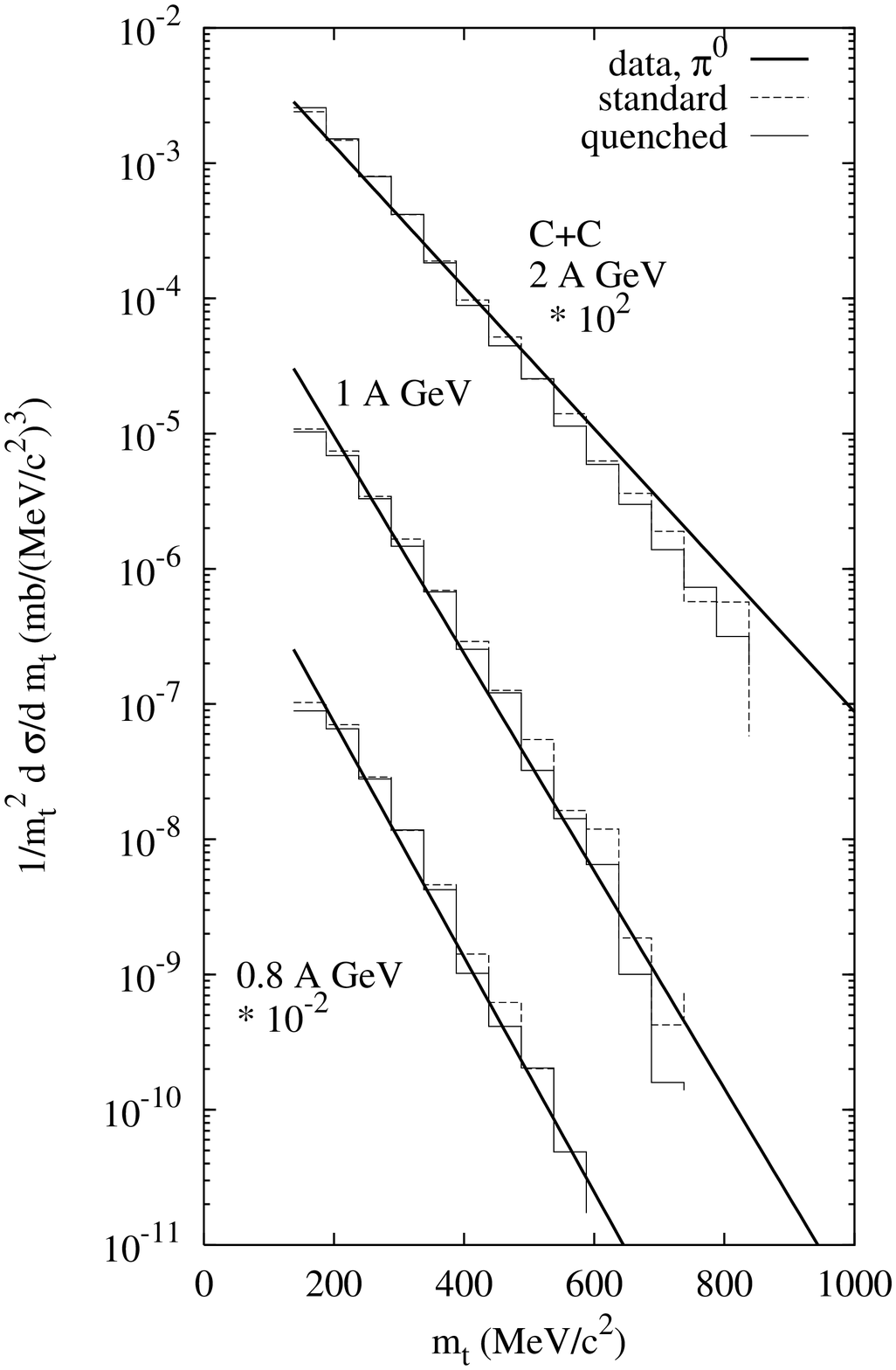,width=0.8\textwidth}

\vspace{2cm}

\caption{ }
\end{figure}

\clearpage
\thispagestyle{empty}

% Fig. 11
\begin{figure}[btp]
\psfig{figure=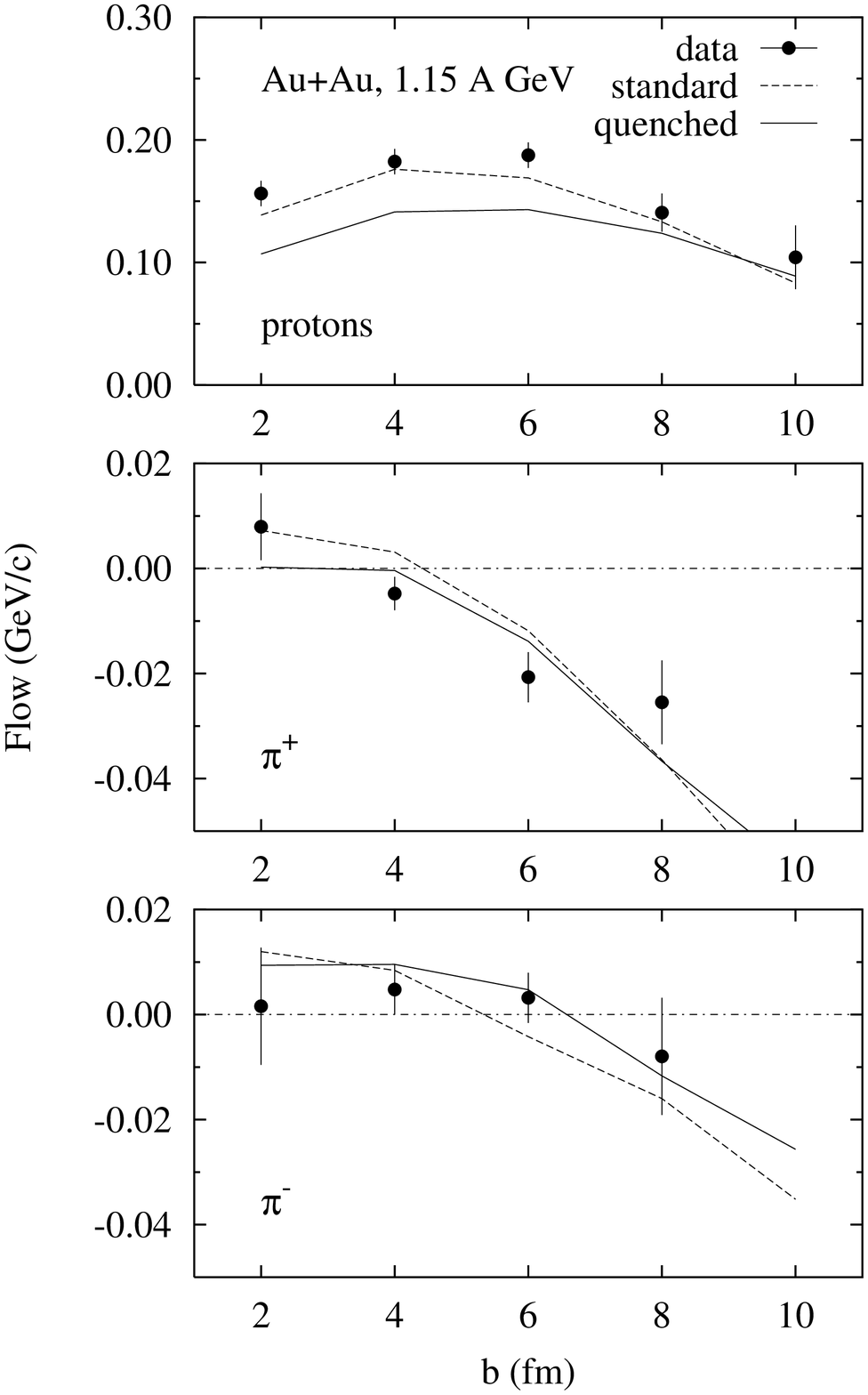,width=0.8\textwidth}

\vspace{2cm}

\caption{ }
\end{figure}

\clearpage
\thispagestyle{empty}

% Fig. 12
\begin{figure}[btp]
\psfig{figure=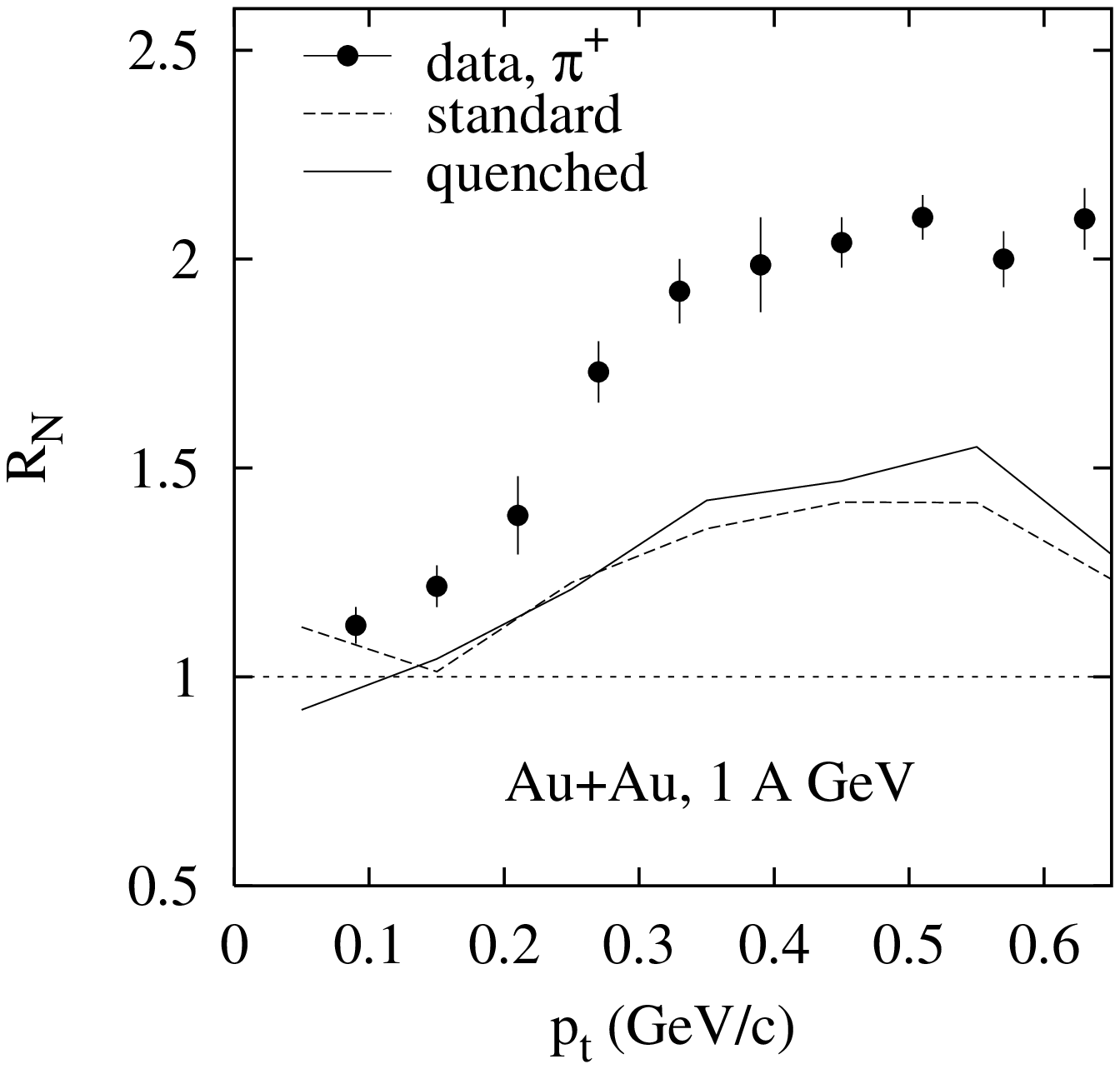,width=0.8\textwidth}

\vspace{2cm}

\caption{ }
\end{figure}


\begin{thebibliography}{99}
\bibitem{Com1} To avoid misunderstanding, we note, that usual Pauli blocking
for nucleons in final states is always included for both standard and quenched
calculations.
\end{thebibliography}
\end{document}